# RXs Directions based Codebook Solution for Passive RIS Beamforming

Ahmed M. Nor
*Telecommunications Department.*
*University Politehnica of Bucharest*
Bucharest, Romania
*Electrical Engineering Department*
*Aswan University*
Aswan, Egypt
ahmed.nor@upb.ro

Seyed Salar Sefati
*Telecommunications Department.*
*University Politehnica of Bucharest*
Bucharest, Romania
sefati.seyedsalar@upb.ro

Octavian Fartu
*Telecommunications Department.*
*University Politehnica of Bucharest*
Bucharest, Romania
octavian.fratu@upb.ro

Simona Halunga
*Telecommunications Department.*
*University Politehnica of Bucharest*
Bucharest, Romania
simona.halunga@upb.ro

*Abstract*—Recently, reconfigurable intelligent surface (RIS) has immensely been deployed to overcome blockage issue and widen coverage for enabling superior performance 6G networks. Mainly, systems use RIS as an assistant to redirect the transmitter (TX) incident signal towards the receiver (RX) by configuring RIS elements' amplitudes and phase shifts in what is known as passive beamforming (PBF) process. Channel estimation (CE) based PBF schemes achieve optimal performance, but they need high overhead and time consumption, especially with high number of RIS elements. Codebook (CB) based PBF solutions can be alternatives to overcome these issues and obtain sub-optimal performance, by only searching through a limited reflection patterns (RPs) and determining the optimal one based on a predefined metric. However, they consume high power and time relevant to the CB size used for searching. In this work, we propose a direction based PBF (D-PBF) scheme, where we aim to map between the RXs directions and the codebook RPs and store this information in an updated database (DB). Using this DB, if the matching between a coming RX and a particular RP exists, the proposed scheme will directly select this RP to configure the RIS elements, otherwise, it memorizes this codeword for searching. Finally, if the matching failed to determine the PBF vector, searching through the memorized RPs will be done to find the optimal one for the RX, then updating the DB accordingly. After a time period, which depends on the CB size, the DB will converge, and the D-PBF scheme will need no searching to select the optimal RP. Hence, the proposed scheme needs extremely lower overhead, power, and time comparable to the CE and conventional CB based solutions, while obtaining acceptable performance in terms of effective rate.

## I. Introduction

Reconfigurable intelligent surfaces (RISs) attract much interest from communication society [1]. Because they are suitable candidates to overcome link blockage issues that appear with millimeter wave (mmWave) and terahertz (THz) bands to enable beyond 5G networks [1]–[3]. For instance, 20 dBm attenuation can influence on mmWave links due to obstacles [4], whether they are static, e.g., buildings, or dynamic, e.g., human bodies. RISs provide another alternative indirect path between transmitter (TX) and receiver (RX), when the main line of sight (LOS) link between them is blocked. This is obtained by reflecting the incident signal on RIS, which comes from the TX, and redirecting it to the RX [1]. In addition, RISs can extend the network coverage, increase the channel rank, and enhance its statistics. Furthermore, the performance of unmanned aerial vehicles (UAVs) based networks can be improved using RIS in public events and emergency situations' scenarios [5].

However, using RIS in different communication networks comes with a critical challenge which is passive beamforming (PBF) process, i.e., determining the amplitudes and the phase shifts of RIS elements. The reason behind this difficulty is that RIS contains high number of passive reflecting elements, thus no control signals can be exchanged between RIS and RX for establishing transmission. Two different directions can be applied to perform PBF, first, channel estimation (CE) based PBF schemes [6], [7], where number of pilots are needed as overhead to estimate the RIS-RX channel then RIS elements' phase shifts can be configured using optimization. But overhead of these methods is extremely high, especially with high numbers of RIS elements. Moreover, estimating the full channel every time period is a complicated process. Furthermore, PBF designs aim to optimize performance by jointly optimizing RIS elements' phases and active TX antennas, which is a time consuming and complex step [8]. Hence, CE based PBF will not efficiently work in large-scale RISs deployments as it needs long time and high overhead. Secondly, codebook based PBF (CB-PBF) techniques direction [8]–[10]. These schemes contain three stages, codebook (CB) generation, training stage and learning step. First, a suitable reflection pattern (RP) codebook is designed. Then, searching along this CB is performed, and based on feedbacks signals, system learns and determines the optimal RP that can obtain the best quality of service (QoS) for the RX [8]. CB-PBF schemes address the pilots' overhead and optimizing RIS elements' phase shifts issues, where there is no need for sending pilot signals to estimate the full channel or performing optimization to jointly design active/passive beamforming vectors. However, CB-PBF cannot achieve similar performance as the CE based solutions because searching is performed on a set of generated codewords, thus it cannot describe the entire universal space. Moreover, the searching process itself consumes time relevant to the CB size. To improve the performance of CB-PBF solutions, an effective CB, that nearly covers the target area, should be generated. In addition, decreasing the searching time can be achieved by reducing the candidate RPs, that shall be examined in the training stage.

Searching and learning steps in the CB-BPF schemes can be considered as the procedures of directing the RIS, i.e., reflecting the TX incident signal, into the position of the RX. Inspired by that if we can map between a specific codeword, RP, in the predesigned CB and the position of the interested user, searching along entire codebook will not be needed. Hence, system overhead and complexity will be decreased. One methodology to do so is to build a database (DB) containing the CB codewords and the corresponding valuable RXs positioning information (PI). Furthermore, this DB shall be updated over time, hence declining the searching complexity and overhead. It is worth to mention that this idea is inspired by the previous works which proposed positioning based active beamforming (ABF) schemes for mmWave communication [11]–[13], where selected ABF vector, i.e., beam directing, is obtained based on the user equipment PI. ABF is straightforward as a relation between beams' direction and positions can be easily defined, in contrast, to best of our knowledge, no relationship is founded between certain RX position and RIS elements' phase shifts in certain codeword.

In this work, firstly, we will generate a set of random RPs codebook with size $Q$ from all space set, which is the original CB for the first time frame. At the beginning, the proposed direction based PBF (D-PBF) scheme will fully search through all RPs, similar to exhaustive search benchmark scheme, to find the codeword that maximizes the objective metric, assuming maximizing RX received power as the target metric. Through progressing on time, our algorithm maps between the selected RP, from searching on the CB, and the RX direction relative to RIS position, i.e., the RX angle, then stores this information in a database (DB). Before establishing a new link through the RIS, the proposed angle or direction based PBF checks the stored DB whether there is a match between the new RX angle and stored ones with a certain threshold, and if yes, it directly uses the RP that corresponds to the matched angle for PBF vector, otherwise, it memorizes this RP in candidate searching vector for further searching. Hence, when matching step failed to determine the PBF vector, the D-PBF scheme searches along the stored candidate RPs, which are $Q_1$ codewords from the CB. As the DB is updated through time until reaching convergence, $Q_1$ will decrease until reaching zero when full mapping is approached. Thus, the proposed scheme will finally need no searching, i.e., system overhead will be zero, to find out the best RP. Consequently, the required overhead of the system is highly reduced compared to other PBF schemes, and its effective achievable rate remains acceptable though it does not select the optimum RIS configuration.

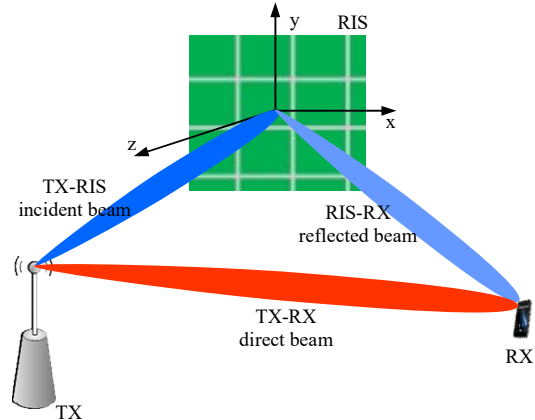

Fig. 1. RIS aided wireless communication system.

The remaining of this paper is constructed as follows: RIS aided wireless communication system model is described in Section II. Then, we review codebook based passive beamforming solutions in Section III. Meanwhile, in Section IV, the proposed angle based passive beamforming technique is presented. Section V shows and discusses the performance of the proposed scheme comparable to benchmarks. Finally, Section VI concludes the paper and suggests future directions.

## II. SYSTEM MODEL

Fig. 1 shows the RIS assisted wireless communication system, where RIS with $N$ passive elements is placed between $M$ element antenna TX and single antenna RX. Here, we assume RIS is used for increasing network coverage as well as overcoming the blockages between TX and RXs, hence both the LOS TX-RX link and reflected indirect reflected TX-RIS-RX link will be existing.

Let $H_o \in \mathbb{C}^{1 \times M}$ denotes the channel from the TX to the RX, $G_1 \in \mathbb{C}^{N \times M}$ refers to the channel between the TX and the RIS, and $G_2 \in \mathbb{C}^{1 \times N}$ refers to the channel between the RIS and the RX. Considering downlink transmission case, the received signal $y_r$ at the RX can be expressed as:

$$y_r = (G_2 \Theta G_1 + H_o) V x_t + n, \quad (1)$$

where $\Theta = diag\{\theta\}$ is the diagonal phase shift matrix of the RIS, here $\theta = [\theta_1, \theta_2, \dots, \theta_N]^T$ is the reflecting beamforming vector at the RIS and $\theta_n = \gamma e^{j\phi_n}$. The $\phi_n \in [0, 2\pi]$ is the phase shift of RIS reflecting element $n$, and $\gamma \in [0,1]$ is the reflection coefficient amplitude. $V \in \mathbb{C}^{M \times 1}$ is the TX beamforming vector, $x_t$ refers to the transmitted symbol by the TX, and $n \sim \mathcal{CN}(0, \sigma^2)$ is the received noise at RX where $\sigma^2$ is the noise power. In RIS assisted communication systems, active-passive beamforming vectors, i.e., $V$ and $\theta$, respectively, should be designed by acquiring the accurate channel state information (CSI). In our study, both TX and RIS are placed in fixed positions, hence the active beamforming vector $V$ can be determined using any appropriate active beamforming scheme [12]. Thus, we will focus only on the PBF process to determine $\theta$ vector.

## III. CODEBOOK BASED PASSIVE BEAMFORMING

CB based PBF solutions require three main stages, first, design the codebook, i.e., determine the set of the RIS RPs, secondly, scan this set and feedback uplink pilots containing information about these RPs, then, find out the optimal RP that optimizes the target metric.

To design the codebook of RPs, the system selects a number of codewords from the universal solution set, i.e., the original codebook which contains all possible RPs. Different types of codebook generation methods can be used, e.g., random CB and sum distance maximization CB [6], [8]. For simplicity, we will consider random CB having $Q$ RPs in our work, where each element phase shifts are randomly selected. Fig. 2 shows an example of random CB with $N = 1$ and $Q = 8$. In the second step, RIS elements are configured according to all $Q$ RPs and feedback for each RP is sent from the user equipment (UE)/RX to the base station (BS)/TX. These feedbacks usually contain the received power at the RX. Using the feedbacks of the $Q$ labeled observations, in a learning step, BS determines the RP for downlink transmission that can obtain the optimum objective function (OF) and maximize target metric (MR). Several learning methods can be utilized, e.g., rote learning, which memorizes all $Q$ observations, then selects the RP that achieves the best OF. This algorithm is simple and straightforward, where its control signal overhead is $\log_2 Q$ bits, so we will consider it in our work. Fig. 3 illustrates this scheme considering the received power at RX as a metric.

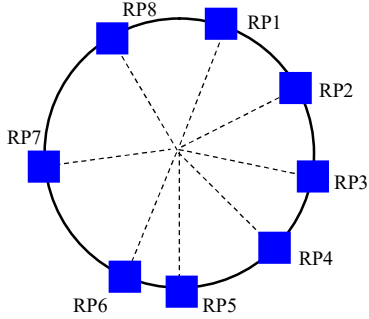

Fig. 2. Random CB generation in CB-PBF schemes.

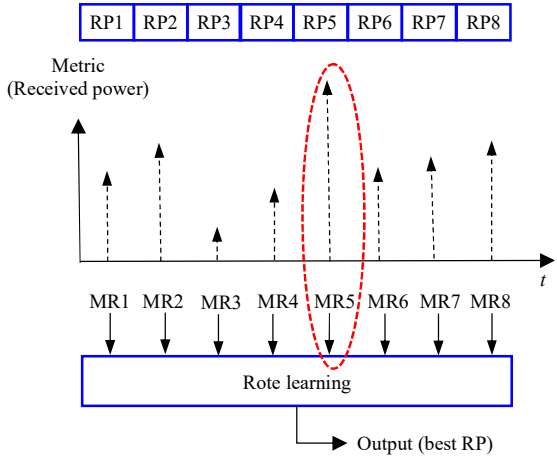

Figure 3. Rote learning in CB-PBF schemes.

## IV. PROPOSED DIRECTION BASED PBF SCHEME

In this section, the description of the proposed direction based passive beamforming (D-PBF) scheme is given. This scheme mainly depends on mapping between the RPs and their corresponding directions of RXs, then storing this mapping information in a database for further usage. Meanwhile, if the proposed scheme failed to determine the passive beamforming vector using the stored DB, a searching and determination phase, followed by a DB updating, is performed. Repeating this procedure over time will obtain convergence to the DB. Consequently, the system can achieve high performance due to extremely lower overhead.

First, the proposed scheme uses the random CB generation with $Q$ RPs, where this number of RPs is determined based on the acceptable beginning complexity. Increasing $Q$ or using more effective and descriptive CB insures better performance at the expense of more computational complexity whether the exhaustive search based PBF, or the proposed schemes are used. Furthermore, the angle threshold, which is a design parameter, is defined as $\varphi_{Thr} = pi/Q$. This threshold controls the database updating speed, the accuracy of mapping between RPs and the directions of RXs, and the accuracy of the selected PBF vector. For instance, larger $\varphi_{Thr}$ quickly stabilizes the database and wrongly maps between RPs and RXs angles. Hence, the system will select improper RP to redirect the incident signal to RX, where this RP will reflect the signal far away from the RX. Regarding the DB, it contains two rows that store the identifications (IDs) of the CB RPs and the corresponding best previous experienced RX direction, $D_\varphi$. This directions' row, $\boldsymbol{D_\varphi}$, is initialized as empty, then it is updated through time until it becomes full. Table I. presents an example of the database at a certain time.

TABLE I. DB STATUS AT CERTAIN TIME.

| RP ID | 1 | 2 | 3 | . | . | $Q-1$ | $Q$ |
|---|---|---|---|---|---|---|---|
| $\boldsymbol{D_\varphi}$ | 40° | 12.5° | ∅ | . | . | 78° | 24° |

When a new RX enters the network, and the TX needs to communicate with this RX with the assistant of the RIS, firstly, the proposed algorithm begins calculating the RX direction relative to the RIS, $\varphi_{RX}$, using the known RIS position and the estimated RX position. Several services, e.g., Wi-Fi, mmWave, visible light communication, can provide the system with a highly accurate RX position with few centimeters errors. Secondly, the searching IDs vector, $\boldsymbol{S_{IDs}}$, and the selecting metric, $\boldsymbol{MR}$, are initialized as empty. $\boldsymbol{S_{IDs}}$ memorizes the RPs IDs that maybe examined, while $\boldsymbol{MR}$ is used to determine and learn the best RP for downlink transmission. Here, we use the received power at the RX as a selecting metric, hence the RP that can maximize it will be utilized for the PBF vector $\boldsymbol{\theta}$. Also, $Q_1$, which indicates to proposed scheme searching overhead, is initialized.

For each $qth$ RP, database and threshold checking step is performed to find out if the RX direction, corresponding to this RP, exists and matches with the current RX angle or not. The matching step is the comparison of the magnitude difference between $\varphi_{RX}$ and $\boldsymbol{D_\varphi}[q]$, and the matching threshold, $\varphi_{Thr}$. The threshold consistency is obtained, if $|\varphi_{RX} - D_\varphi[q]|$ is smaller than or equals to the predefined threshold. Consequently, the matched codeword will be selected to adjust the PBF vector $\boldsymbol{\theta}$, and one element in the database will be filled with the current RX angle. Meanwhile, if matching fails for this $qth$ RP, searching IDs vector will memorize this RP ID for further searching and $Q_1$ will increase. Through progressing on time, as new RXs require communication, system keeps updating DB with new matching pairs, i.e., $qth$ RP and $\varphi_{RX}$, until DB converges after $T$ frames. This convergence required time depends on the initial CB size $Q$ as we will illustrate.

In case of matching failure between the current RX angle and all stored directions, system will perform the searching phase. In this stage, system examines the RPs IDs, that are stored in $S_{IDs}$, and configure the RIS using these codewords. Thereafter, for each $q_1th$ RP, the searching metric, $MR[q_1]$, will be calculated and stored. Then, the proposed scheme selects the RP, that guarantees the maximum searching metric, $MR$, for configuring RIS using it, and updates database accordingly. Consequently, the system overhead for a certain RX, if system needs the searching phase, is $\mathcal{O}(Q_1)$, where $0 \leq Q_1 \leq Q$. Furthermore, the complexity for RIS configuration is $\mathcal{O}(N)$, if matching happens, while it is $\mathcal{O}(Q_1 N)$, otherwise. It is worth to mention that empty database means $Q_1 = Q$, and the D-PBF scheme searching complexity will equal exhaustive search one. But $Q_1$ decreases through time until equaling one when DB converges, where the PBF will be done without searching, i.e., just using angles' matching. Algorithm 1 summarizes all the steps of the proposed D-PBF scheme. The worst case total computational complexity of the proposed scheme, when DB converges, is $\mathcal{O}(3Q + N)$, which represents step 5-7 in algorithm 1, where they will be the only required steps for configuring the RIS elements. Furthermore, the system overhead will be $\mathcal{O}(1)$ when using our proposed scheme after DB convergence.

## V. SIMULATION RESULTS

In this section, the required overhead when using the proposed angle based PBF scheme is presented over time. Moreover, the proposed scheme performance is compared with the three benchmarks, the CE based PBF, the full random CB based PBF, and random configuration schemes, in terms of the effective achievable rate, $R_{eff} = (1 - \tau/T)R$, where $\tau$ and $T$ are the searching overhead and channel coherence time, respectively, while $R$ represents the achievable rate. The CE-PBF scheme optimizes the PBF vector based on alternating optimization (AO) algorithm, meanwhile, the CB-PBF examines all RPs to find out the best one that can be used. Moreover, we will discuss the required complexity of the D-PB comparable to other schemes. In the following subsections, we will show our simulation scenario and results.

### A. Simulation Scenario and parameters

In this study, RIS aided a multi input single output (MISO) system is considered. We use the same channel parameters as mentioned in [7] unless we differently define

**Algorithm 1:** The proposed D-PBF scheme.

**Inputs**: The codebook $\boldsymbol{C}$ with $Q$ RPs, and the angle threshold, $\varphi_{Thr}$
**Initialization**: $\boldsymbol{D}_\varphi = \emptyset, \boldsymbol{\theta} = \emptyset$.
1. **for** each new RX **do**
2.    **Calculate** the RX direction relative to RIS, $\varphi_{RX}$
3.    **Initialize** $S_{IDs} = \emptyset, MR = \emptyset, Q_1 = 0$
4.    **for** $q = 1$ to $Q$ **do**
5.      **if** $\boldsymbol{D}_\varphi[q]$ is exist **and** $|\varphi_{RX} - \boldsymbol{D}_\varphi[q]| \leq \varphi_{Thr}$ **then**
6.        $\boldsymbol{\theta} = \boldsymbol{C}[q], \boldsymbol{D}_\varphi[q] = \varphi_{RX}$,
7.        **return**
8.      **else**
9.        $Q_1 +$,
10.       $S_{IDs}[Q_1] = q$,
11.     **end if**
12.   **end for**
13.   **for** $q_1 = 1$ to $Q_1$ **do**
14.    **Configure** RIS elements using $\boldsymbol{C}[S_{IDs}[q_1]]$.
15.    **Calculate** and **memorize** $MR[q_1]$
16.   **end**
17.   **Find** $q^* = arg\,max\,MR$
18.   $\boldsymbol{\theta} = \boldsymbol{C}[S_{IDs}[q^*]], \boldsymbol{D}_\varphi[S_{IDs}[q^*]] = \varphi_{RX}$
19. **end for**
**Output**: The passive BF vector $\boldsymbol{\theta}$.

TABLE II. SIMULATION PARAMETERS.

| Parameter | Value |
| --- | --- |
| TX signal power | 0.001 Watt |
| TX location | (18, 24, 50) |
| RIS location | (0, 0, 50) |
| RXs height | 1m |
| Number of TX antennas | 16 |
| Number of RIS elements | 16 |
| Noise power spectrum density | -160 dBm/Hz |
| Channel coherence time, $T$ | 500 |
| System bandwidth | 10 MHz |

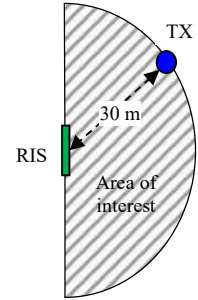

Fig. 4. Simulation area.

one. Table 2 summarizes the simulation parameters, where an area of interest, in the shape of half circle with a diameter 30$m$, surrounded the RIS is considered. This target area is shown in Fig. .4. Furthermore, we assume uniformly distributing RXs in this area, then, to focus only on angle effect, we consider RXs to be distributed on the edge of the target area circle as a second simulation scenario. Moreover, for the designed codebook, we use a random CB generation for both the conventional CB scheme and the initialization of the proposed angle based CB. The results are measured by averaging over 100 realizations.

*B. Results*

Fig. 5. presents the proposed angle based passive beamforming scheme searching overhead, $\mathcal{O}(Q_1)$, versus the progress in time, when RXs are distributed through the area of interest. The required overhead exponentially decreases through time taking into consideration the initial codebook size. Hence, when the database converges, the proposed scheme needs no searching overhead to select the best RP, i.e., $Q_1 = 0$, and a $Q$ unused time slots will be available for data transmission (DT) in additional to the pre-allocated slots. For example, with CB size equals to 100 and 200, the proposed scheme overhead reaches zero after 1000 and 1920 time frames, respectively. On contrast, the conventional CB-PBF overhead will remain $\mathcal{O}(Q)$, where $Q$ equals 100 and 200, respectively, meanwhile CE-PBF scheme overhead is $\mathcal{O}(N)$. Hence, using the proposed D-PBF scheme, a very high reduction in overhead can be obtained. Distributing RXs over the area is not fair for the proposed scheme which uses only the angle in database updating, while the distance between RIS and RX effects on the performance. Thus, we study the searching overhead of it, if the second scenario is applied, in Fig. 6. In this case, with CB size equals to 100 and 200, the proposed scheme overhead reaches zero after 945 and 1625 frames, respectively. This means that the proposed scheme was more able to map between the RXs directions and the predefined RPs, and update DB quicker than the first case. Because, in this scenario, the angle is the only positioning parameter that effects on the system, as the distance was neutralized.

Fig. .7 shows the proposed scheme performance in terms of effective achievable rate in bps/Hz when the database converges versus CB size and comparable to benchmarks algorithms. CE-PBF outperforms all schemes, however it needs impractical requirements such as high computational complexity, $\mathcal{O}(N_i N^{4.5})$, for optimizing the RIS elements, where $N_i$ is the number of iteration in AO method. Moreover, CE overhead is $\mathcal{O}(N)$. Increasing the codebook size, $Q$, to a value larger than the number of RIS elements in conventional CB scheme is also impractical, because, in this case, system will require higher overhead than that needed by CE-PBF, causing decreasing in effective achievable rate. On contrast, the proposed scheme enhances the obtained effective rate with the increasing of CB size, as no channel estimation is required at all. For example, using 150 RPs CB, the system achieves 5.99 and 6.18 bps/Hz effective rate in first and second scenario, respectively, which are only 23% and 18% lower than that obtained by CE-PBF scheme, respectively. Moreover, it is noted that no further improvements in the performance of the proposed scheme can be achieved after a certain CB size, which is nearly 150. Although the proposed D-PBF scheme obtains lower performance than CE-PBF, it needs extremely lower computational complexity than AO procedure to determine RIS configuration after reaching DB convergence. Fig. 8 shows the required computational complexity of the proposed scheme comparable to other benchmarks versus different CB sizes. Here, we consider $N_i$ equals 3, but AO

will need more iteration in real environments to converge the RIS elements' phase shifts. Using 150 RPs CB, the proposed D-PBF scheme needs nearly $1.6 \times 10^3$-times lower complexity than CE-PBF scheme.

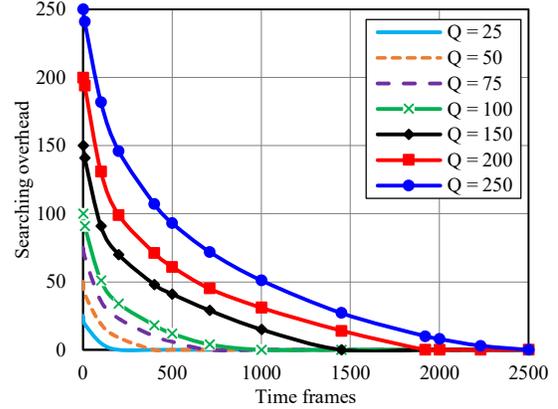

Fig. 5. The D-PBF scheme searching overhead versus progress in time, when applying first scenario.

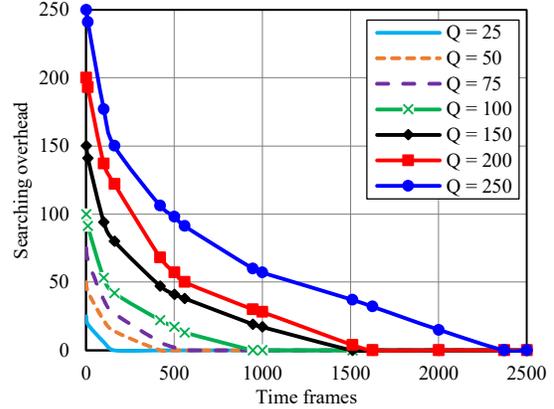

Fig. 6. The D-PBF scheme searching overhead versus progress in time, when applying second scenario.

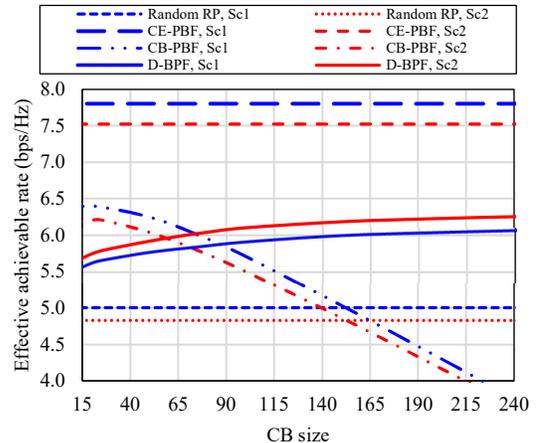

Fig. 7. The effective achievable rate of the proposed scheme compared with other benchmarks.

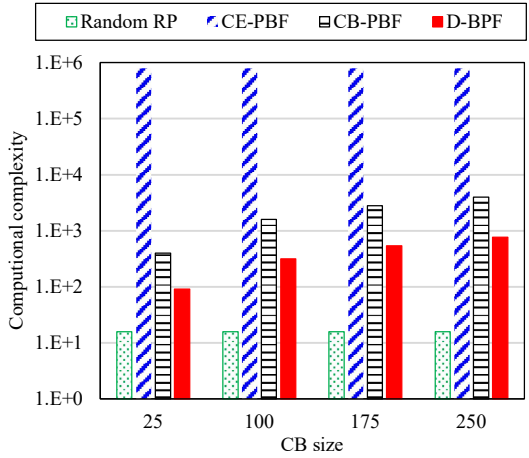

Fig. 8. The computational complexity of the proposed scheme versus different CB sizes.

## VI. CONCLUSION

A novel context information based passive beamforming scheme is presented in this paper, where a DB, containing the RPs and their corresponding RXs directions, is built and updated over time. consequently, for each new RX, the proposed D-PBF tries to match between the new RX angle and previous stored directions, to select the RP that achieves angles matching within threshold, and otherwise, it examines along the codewords, that are chosen as candidates for searching, and selects the RP that maximizes the objective function. As a result of DB convergence after a number of time frames, the required overhead of the system is highly reduced comparable to other PBF schemes. Hence, its effective achievable rate remains acceptable though it does not select the optimum RIS configuration. This work can be extended by considering other context information parameters, e.g., RXs positions and orientations, and updating DB based on additional conditions. Furthermore, using more efficient codebook generations methods will guarantee better performance.

ACKNOWLEDGMENT

This study has been conducted under the project 'MObility and Training fOR beyond 5G eco-systems (MOTOR5G)'. This project has funded from the European Union's Horizon 2020 programme under the Marie Skłodowska Curie Actions (MSCA) Innovative Training Network (ITN) under grant agreement No. 861219.